# Dynamics of trapped magnetic flux in superconducting $FeTe_{0.65}Se_{0.35}$

S.I. Bondarenko[1], A.N. Bludov[1], V.P. Koverya[1], S.I. Link[1], A.G. Sivakov[1], V.P. Timofeev[1], D.J. Gawryluk[2,3], R. Puźniak[2] and A. Wiśniewski[2]

[1] *B. Verkin Institute for Low Temperature Physics and Engineering, National Academy of Sciences of Ukraine, pr. Nauky 47, 61103 Kharkov, Ukraine*
[2] *Institute of Physics, Polish Academy of Sciences, Aleja Lotników 32/46, PL-02668 Warsaw, Poland*
[3] *Laboratory for Scientific Developments and Novel Materials, Paul Scherrer Institute, CH-5232 Villigen PSI, Switzerland*

e-mail: bondarenko@ilt.kharkov.ua

**Abstract**

The magnetic moment in the superconducting and normal state of a crystalline $FeTe_{0.65}Se_{0.35}$ superconductor, grown by the Bridgman's method with relatively high growth rate, was measured. The temperature and magnetic field dependences of magnetization and its relaxation time were determined. Studied crystal, being non-uniform due to high growth rate of 5 mm/h, exhibits smaller width of superconducting transition in comparison with an ideal crystal grown with velocity of 1 mm/h, and the difference in magnetic properties of crystals grown with various growth rate, related to their microstructure, is discussed.

## 1. Introduction

Experimental studies of the dynamics of magnetic flux in high-temperature superconductors (HTSC's) are important for understanding the basic mechanisms of pinning and creep of Abrikosov and Josephson vortices and for development of a rigorous theory of this process. The behavior and characteristics of flux composed of weakly interacting vortices emerging in a small dc magnetic field of the order of oersted up to several hundred oersted are seldom studied [1, 2].

Magnetization (*M*) of the second type superconductor, caused by induced diamagnetic currents or by the trapped magnetic field (TMF), relaxes with time [3]. This process is especially noticeable for high-temperature superconductors. TMF relaxation, due to thermally activated creep of vortices and jumps of vortices, depends on the effectiveness of pinning centers. The pinning centers are the natural and artificial structural defects within the sample where the superconducting order parameter is suppressed and free energy of superconductors with TMF has a local minimum. Probability of vortex motion increases exponentially with increasing temperature ($T$) and is inversely proportional to the effective depth of the energy potential ($U$) of pinning centers. Microstructure of high-temperature superconductors and the corresponding value of $U$ play a crucial role in the dynamics of magnetic flux [4, 5]. In the simplest case of a linear Kim-Anderson model [1, 2], a value of $U$ can be estimated from measurements of the normalized rate of isothermal magnetization relaxation ($S$) of superconducting sample:

$$S = 1/M_0\,(dM/d\ln t) = -\,k_B T/U, \qquad (1)$$

where: $M_0$ – initial value of the superconductor magnetization ($t = 0$), $t$ – time, $k_B$ – Boltzman's constant.

This paper presents the results of experimental study of the dynamics of the trapped small density magnetic flux in a sample of iron-based superconductor $FeTe_{0.65}Se_{0.35}$ (further abbreviated in the text as FeTeSe) belonging to the family of chalcogenides. In contrast to our previously study of magnetic properties of single crystal of this compound [6] grown by the Bridgman's method with a low growth rate (1 mm/h), this paper reports flux dynamics in the crystal of the same chemical composition, but grown with a high growth rate (5 mm/h). Some features of the structure of the crystal grown at a high rate are clearly visible in Fig. 1.

The sample consists of aligned layered single micro-crystallites with the sizes of tens of microns with slightly disorientated planes. The X-ray study has shown that the sample contains tetragonal phase with some inclusions of hexagonal phase. Further in the text micro-crystallites planes are denoted as the *a-b* planes and the direction perpendicular to this plane is denoted as the *c* direction.

Measurements of temperature dependences of magnetic moment $m(T)$ of the sample for various values of dc magnetic field applied in two mutually perpendicular orientations ($H \| a\text{-}b$ and $H \| c$) were carried out. For several values of temperature near the critical temperature $T_c$ isothermal magnetization relaxation $M(t)$ was studied. The value of normalized rate of isothermal relaxation $S$ was calculated and its dependences on $T$ and $H$ were established. Finally, estimations of averaged values of the effective pinning potential $U$ were made using eq. (1). For crystals of FeTeSe, grown with different growth rate, magnetic characteristics were compared. In addition, measurements of the magnetization hysteresis loops $m(H)$ at various temperatures for different field orientations, with respect to the *a-b* plane (up to 5000 Oe) were carried out. The estimation of the critical current density ($j_c$) and its temperature dependence was also performed.

## 2. Experimental details

The sample with dimensions of 2.5×1.9 $mm^2$ in the *a-b* plane, and with a thickness of ~ 0.5 mm along the *c*-axis, was cut out from the bulk $FeTe_{0.65}Se_{0.35}$ crystal using the laser beam technique (Fig.1a). The magnetic moment of the sample was measured using Quantum Design SQUID magnetometer MPMS-5 in dc magnetic fields $0 \leq H \leq 5000$ Oe in the temperature range 5 K $\leq T \leq$ 300 K. To exclude undesirable and parasitic TMF, all cycles of measurement were started after sample heating to the temperature $T >> T_c$ .

## 3. Results and discussion

Temperature dependence of magnetization $M = m/V$ ($V$ – crystal volume) normalized to the maximum diamagnetic value $M_{max}$ for $H \| c$ is shown in Fig. 2. The temperature dependence of $M/M_{max}$ for $H \| a,b$ is qualitatively similar. These data allow us to estimate critical temperature of a superconductor ($T_c^{onset}$ = 14 K). High growth rate of the crystal has not affected the value of $T_c^{onset}$ [6]. Small width of the transition indicates high degree of structural uniformity of the crystal. Temperature dependence of the magnetic moment $m$ ($T$) for $H \| c$ in a wide range of temperatures is shown in the inset. The feature is observed at $T$=120 K which can be connected with Verwey transition in these systems. In particular, existence of Verwey transition in magnetite $Fe_3O_4$ at $T \approx 120$ K is a fact of common occurrence [7]. X-ray studies have shown that the sample contains magnetite ($Fe_3O_4$) impurity (less than 0.3 per cent of the sample mass), which can be responsible for the appearance of the feature observed in Fig. 2 at $T = 120$ K [7].

Isothermal relaxations of sample magnetization $M(t)$ at $T$ = 5 K, normalized to initial value of $M_0$ at $t$ = 0, for two orientations of magnetic field ($H$ = 50 Oe), are shown in Fig. 3,a. $M(t)$ dependences can be approximated well by the logarithmic law (Fig. 3,b) as it was predicted by the thermally activated creep linear model [1, 4, 5].

Obtained $M(t)$ data allowed us to determine normalized isothermal relaxation $S$, and its dependence on temperature and magnetic field. Temperature dependences $S(T)$ for two orientations of magnetic field $H$ = 50 Oe are shown in Fig. 4,a. One can see that the value of $S(T)$ is increased as $T$ approaches $T_c$. The layered microstructure of a crystal leads to appreciable anisotropy of dynamic properties of magnetic flux in FeTeSe crystal. For $H \| a$-$b$ the vortexes cores are aligned in the *a-b* planes and effectively pinned by planes containing Fe ions [6, 7]**.** The free energy of a superconductor for such configuration of vortices is minimized. As a result, $S(T)$ dependence at $H \| a$-$b$ is found to be below the dependence $S(T)$ at $H \| c$. Thus, $S(T)$ dependences shown in Fig. 4,a confirm the assumed physical model. Observed increase of the $S$ value near $T_c$ points out the mechanism of thermally activated flux creep and is in agreement with experimental data for cuprate HTSC's [1, 2, 10].

Data presented in Fig. 4,b show effective pinning potential $U(T)$ for the crystal grown with high (5 mm/h) and low (1 mm/h) speed, respectively. In the first case, much bigger number of the effective pinning centers is expected (both in the layers of micro crystallines and in area of interlayer multiconnected contacts between them [8]). Apparently, this leads to increase of effective pinning potential by more than one order of magnitude, as is observed in our experiment (Fig. 4 b).

According to our knowledge, there are no reports on $S(H)$ dependence for Fe-based superconductors. For cuprate superconductors it was shown [1] that relaxation rate $S(H)$ is proportional to $H^3$ for magnetic fields $H \leq 20$ kOe. Such dependence was explained assuming only a partial penetration of magnetic flux into superconductor [1]. Another explanation of cubic-power-law dependence assumes possible influence of the surface barriers. In Fig. 5 our experimental data for $S(H)$ dependence for $H \parallel c$ in FeTeSe are shown. The field dependence of $S(H)$ for $H \parallel a\text{-}b$ is qualitatively similar. One may conclude that $S(H)$ dependence is close to $H^{0.7}$ for $5 \leq H \leq 500$ Oe. Such behavior of the relaxation rate can be related to an increase of the effective pinning potential $U$, as it was observed earlier in cuprate HTSC's for low dc magnetic fields [1, 2]. Similar behavior for our FeTeSe superconductors is show in inset to Fig. 5, where $U(H)$ dependence is presented in a double logarithmic scale. Additional study is required for the explanation of found dependence.

The width of a magnetic hysteresis loop, $\Delta M$ at $H = const$, is proportional to effective pinning potential $U$ averaged over the sample volume. The value of a critical current density $j_c$ is determined by a geometrical shape of the investigated sample and $\Delta M$ value according to Bean model of the critical state [9, 10]. For the sample with rectangular cross-section, $j_c$ can be calculated according to the formula:

$$j_c = 20\,\Delta M/[a(1 - a/3b)], \quad (2)$$

where $a, b$ ($a < b$) are the sizes of the sample cross-section.

Magnetization loops $m(H)$ of the investigated sample at two temperatures for $H \parallel c$ are shown in Fig. 6,a. The magnetization loops are not fully symmetric with respect to $H$-axis. This may be caused by possible influence of the Bean-Levingston surface barriers as well as the presence of not compensated magnetic moment of Fe ions in the sample [6]. Temperature dependences of $j_c$, calculated using eq. (2) for $H = 0$ and $H$ =2000 Oe are presented in Fig. 6,b ($H \parallel c$). The data show that high critical current density up to $10^5$ A/cm$^2$ can be achieved in tested FeTeSe superconductor at $H$ =2000 Oe. This is important for practical application of this type of Fe-based superconductors.

## 4. Conclusions

The effective pinning potential $U(T)$ for the $FeTe_{0,65}Se_{0,35}$ crystal grown with high speed (5 mm/h) is more than order of magnitude larger than that observed for the crystal grown with small speed (1 mm/h). This difference is caused by much larger number of the effective pinning centers occur in the crystal grown with higher speed.

In a wide time interval a magnetic relaxation of trapped magnetic flux of small density is described well by the linear model of thermally activated creep. It was shown that magnetic field dependence of isothermal relaxation of magnetization $S(H)$ in $FeTe_{0,65}Se_{0,35}$ superconductor grown up with high rate (hence possessing many effective pinning centers), in a range of low magnetic fields ($H < 1000$ Oe) is close to $H^{0.7}$.

Large values of pinning potential and the critical current density determined from magnetic measurements agree very well with results obtained from transport measurements [8].

**Acknowledgement** This work was partially supported by the National Science Centre of Poland based on decision No. DEC-2013/08/M/ST3/00927.

## Figures

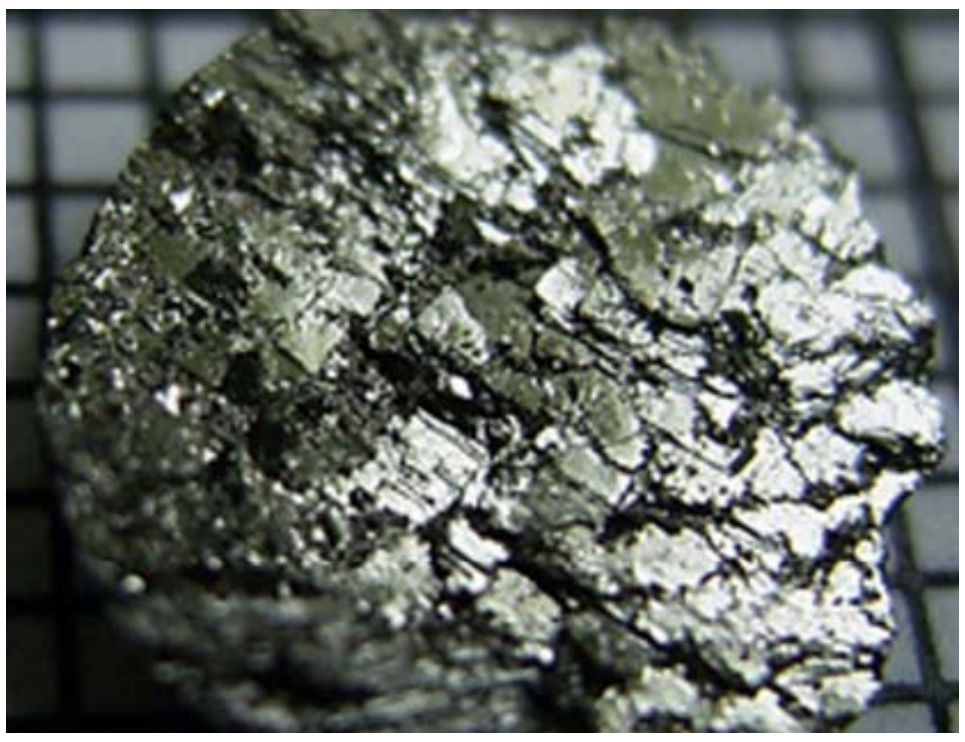

a)

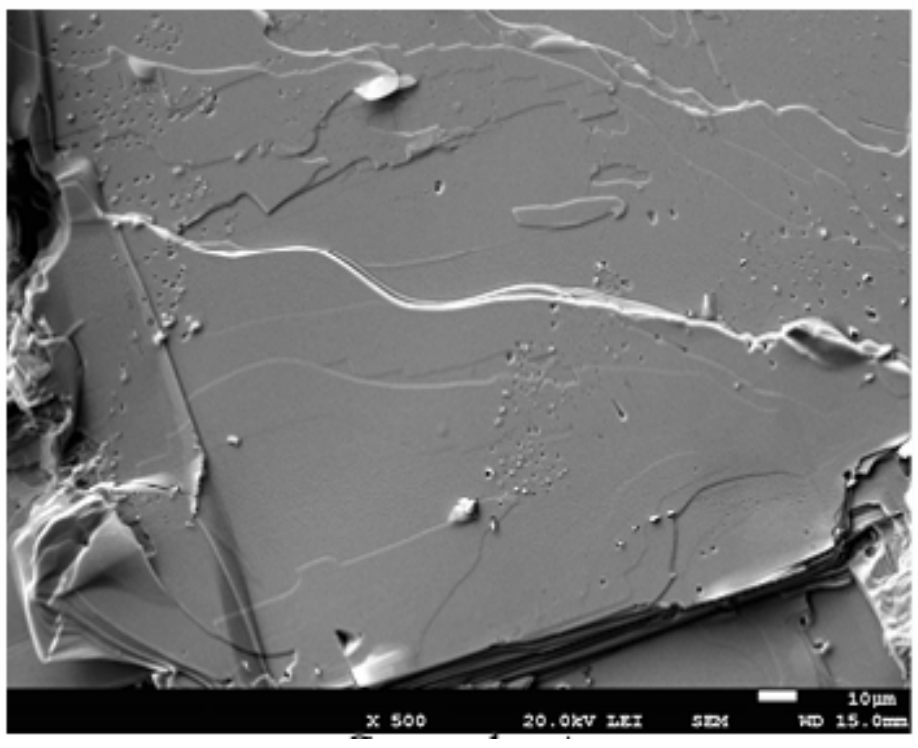

b)

Fig.1 a) External view of a crystal grown at a high rate (the grid step corresponds to 1mm).
b) Field emission scanning electron microscopy image of the surface of the sample grown at a rate 5 mm/h. The (001) crystal plane is parallel to the surface.

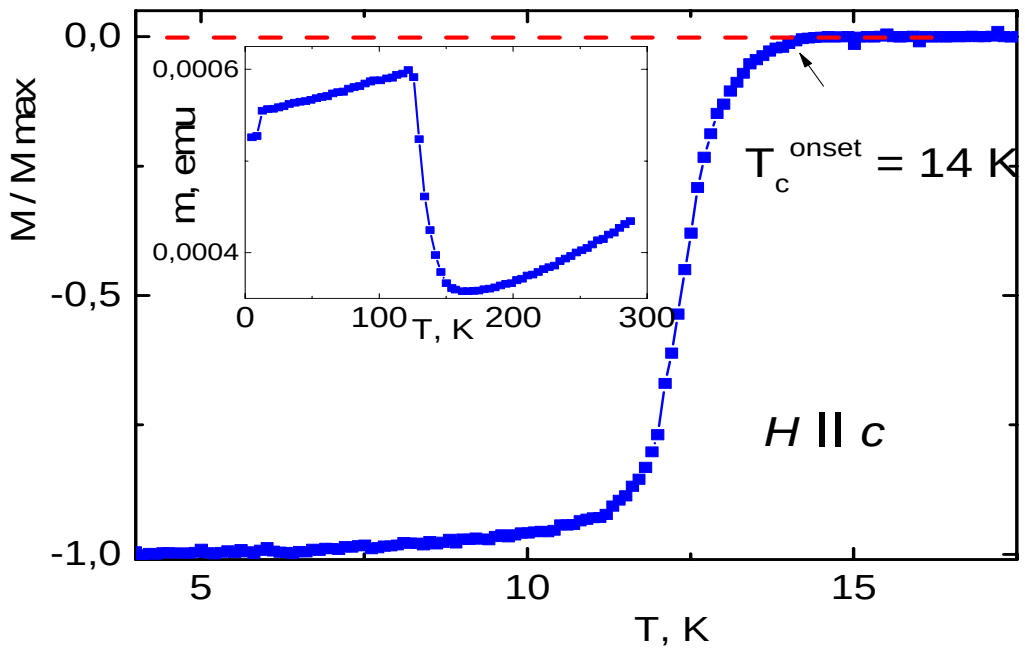

Fig. 2. Temperature dependence of relative magnetization of the sample normalized to maximum diamagnetic value $M_{max}$ (positive magnetic moment of excess Fe and/or $Fe_3O_4$ has been deducted) at $H \| c$. Inset: temperature dependence of the magnetic moment in a wide temperature range.

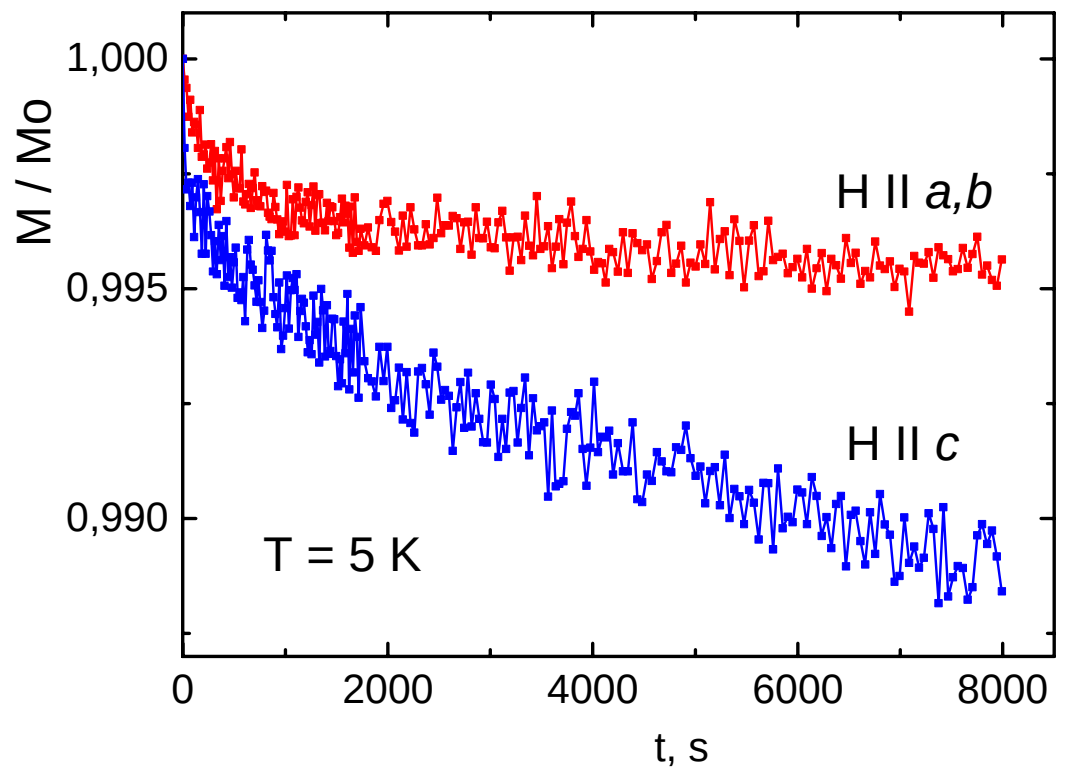

a)

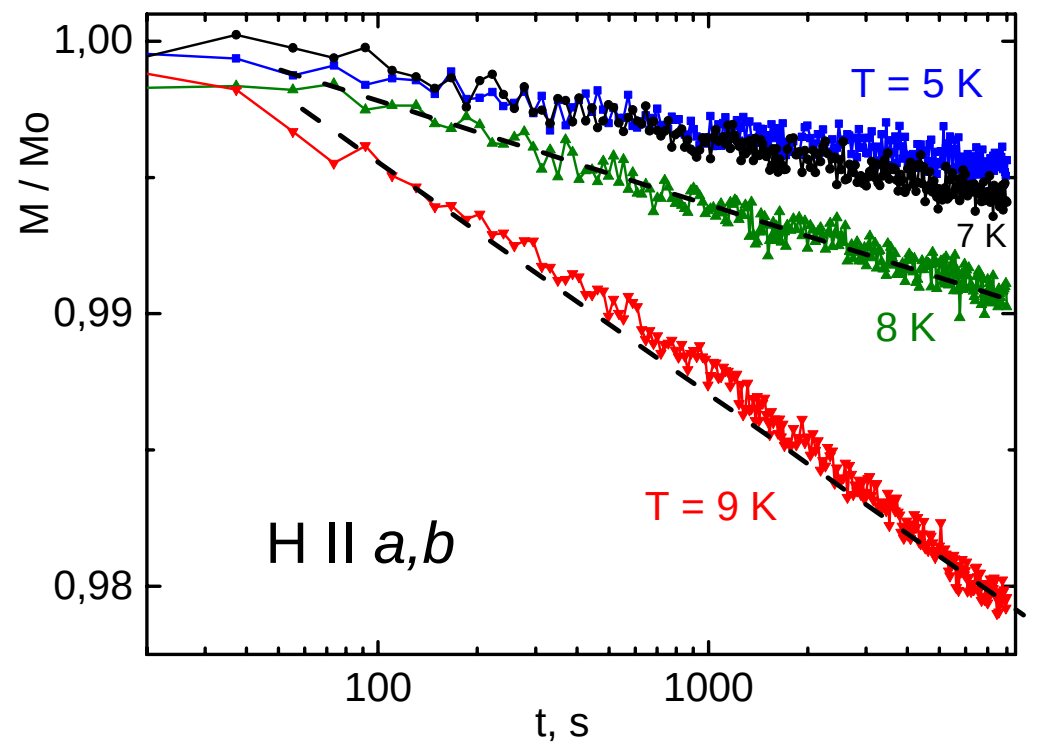

b)

Fig. 3. a) Normalized isothermal relaxation of magnetization $M(t)/M_0$ at $T$ =5 K and constant magnetic field $H$ = 50 Oe for two orientations of magnetic field. b) The $M(t)/M_0$ dependences for $H \parallel a,b$ in semilogarithmic scale. Dashed straight lines demonstrate $M(t)/M_0 \sim \ln t$ dependences.

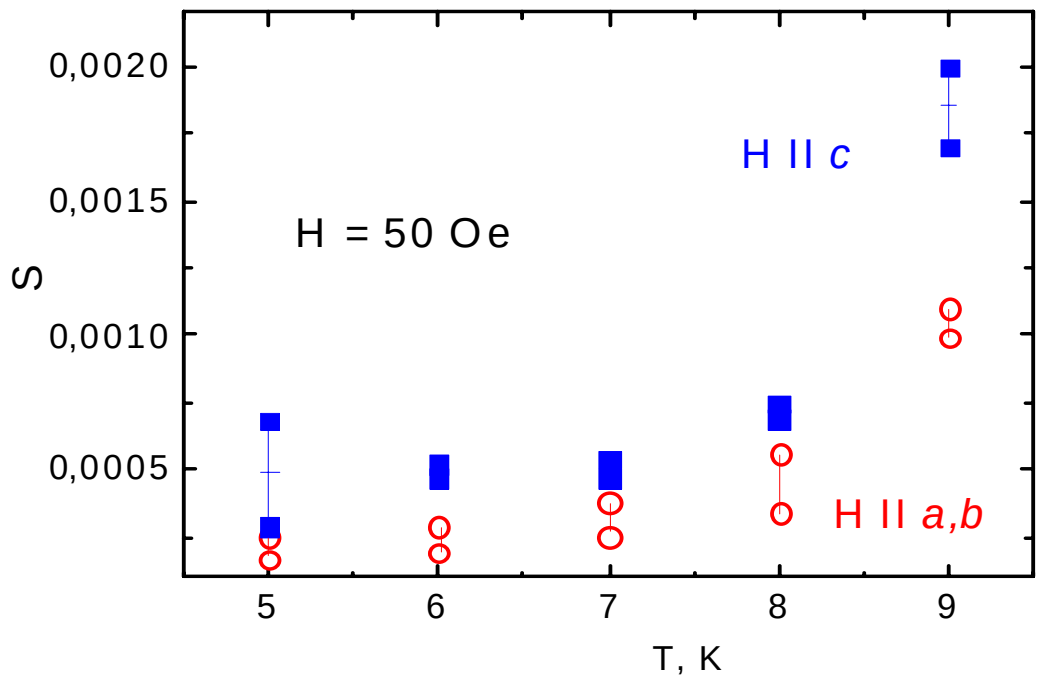

a)

b)

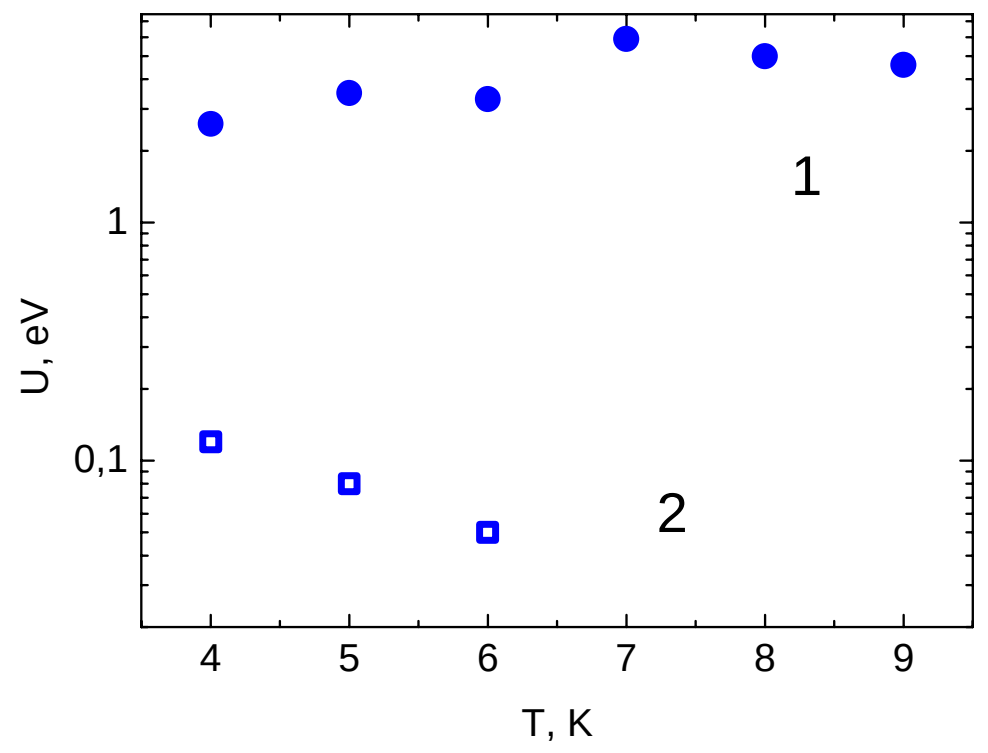

Fig. 4. a) Temperature dependences of the isothermal relaxation of magnetization $S$ for two orientations (squares – $H \parallel c$, circles – $H \parallel a,b$) of magnetic field $H$ = 50 Oe . Scattering of experimental data illustrate top and bottom symbols at each temperature. b) Effective pinning potential $U(T)$ for the crystal grown with high speed (1) and with low speed (2) for $H \parallel c$.

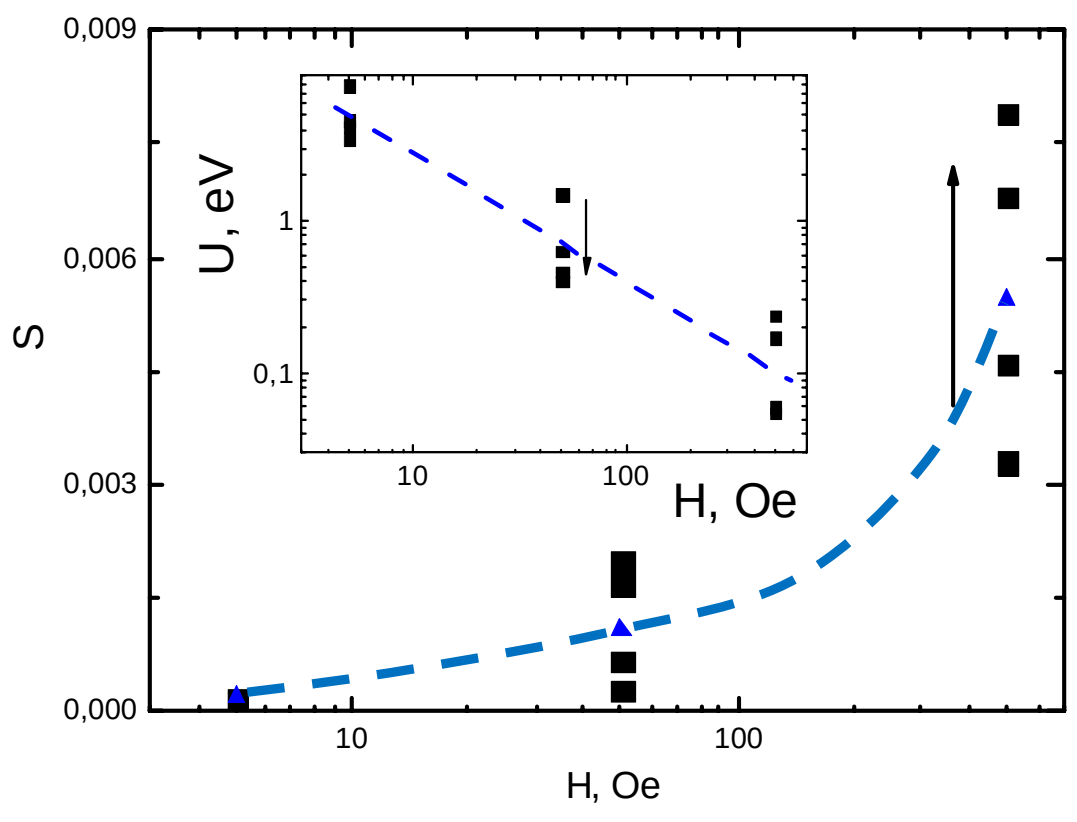

Fig. 5. Field dependence of isothermal relaxation of magnetization $S$ for $H \parallel c$ in the temperature range of 5 – 9 K (arrow indicates increasing temperature). The dashed line illustrates $S \sim H^{0,7}$ dependence. Inset: field dependence of $U$ in double logarithmic scale. Dashed lines are the guidance for eye.

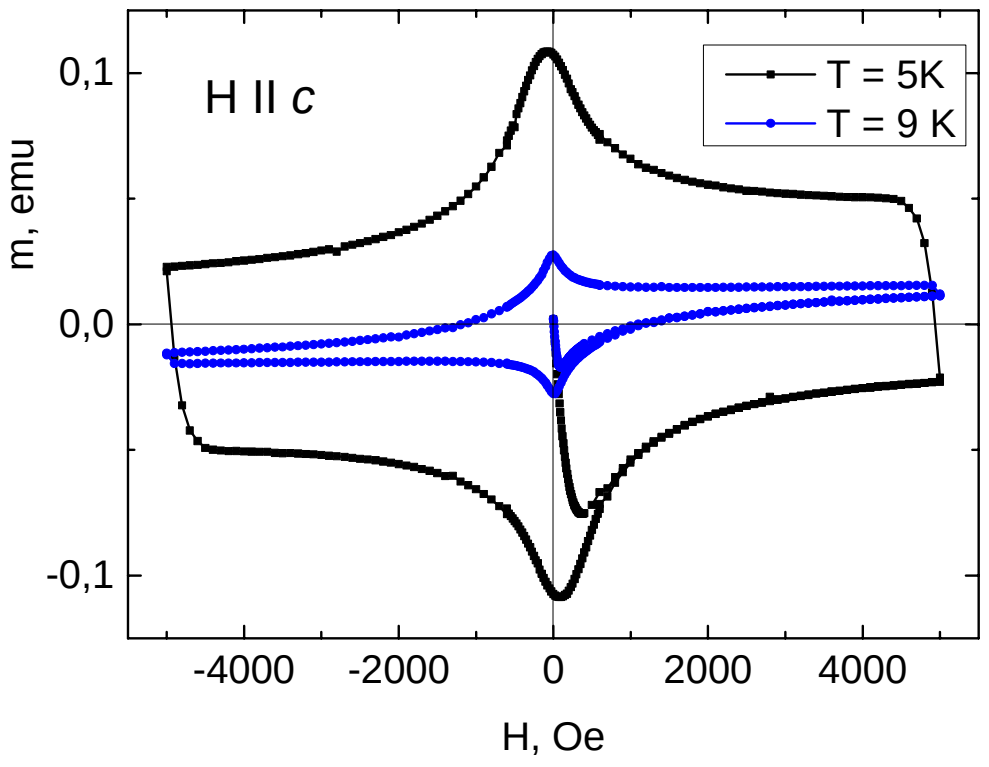

a)

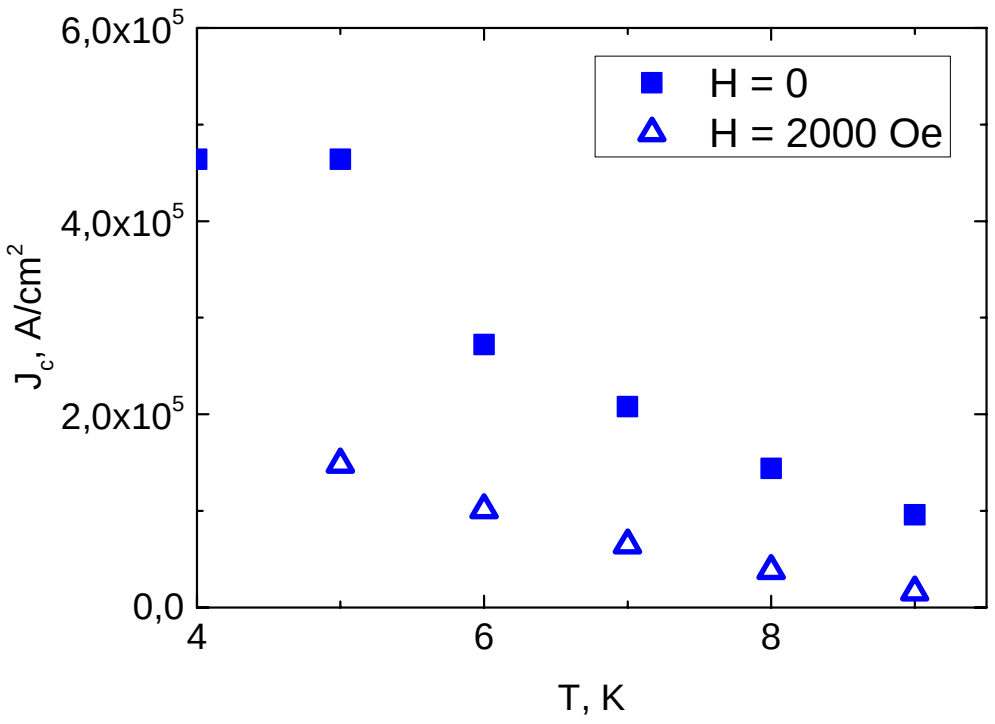

b)

Fig. 6. a) Hysteresis loops $m(H)$ at two temperatures for $H \| c$; b) temperature dependences of $j_c$, calculated using eq. (2), for $H \| c$ (value $\Delta M$ was taken at $H = 0$ and 2000 Oe).